\documentclass{an}
\usepackage{graphicx}
\usepackage{times}
\usepackage{fancyhdr}
\begin{document}
\include{inc_header}

\title{For whom the disc tolls}

\author{J.-P. Lasota\inst{1,2}}
\institute{Institut d'Astrophysique de Paris UMR 7095 CNRS \and
Universit\'e Pierre \& Marie Curie 98bis Boulevard Arago, 75014 Paris, France}

\date{Received 8 September; accepted 22 September 2005;
published online 20 October 2005}

\abstract{
Report on the Nordita
Workdays on Quasi-Peridic Oscillations (QPOs).
}

\correspondence{lasota@iap.fr}

\maketitle


The publications  resulting from the   Nordita Workdays on QPOs
are an interesting and original contribution to research on
accretion flows around compact objects.
They contain four observational papers, one theoretical
paper dealing with numerical simulations of accretion discs and
eleven contributions (some of them analyzing observations) totally
devoted to the epicyclic resonance model (ERM) of high frequency
QPOs (hfQPOs) of Abramowicz \& Klu\'zniak. Probably all that is to
be known about this model is included in these publications. This is
their strength but also their weakness. First the model is not
complete, it is rather kinematic than dynamic. It describes in great
detail the interactions between two oscillations but as Klu\'zniak
confesses: \textsl{It would be good to identify the two non-linear
oscillators.} Yes indeed. Not only \textsl{good} but crucial.
Second, concentrating on hfQPOs only is most probably not a wise
decision because there exist (admittedly complex) relations between
them and their lower frequency brethren and there is a clear link
between their presence and the state of the system. Although the
authors of the eleven papers sometimes pay lip-service to
observations not directly concerning the frequency values of hfQPOs,
in practice they seem to ignore the very important conclusion of
Remillard: \textsl{... models for explaining hfQPO frequencies
must also explain the geometry, energetics and radiation mechanisms
for the SPL state}. By the way, probably even this will not do: the
model will have to explain all the X-ray states. One can understand
the reluctance to leave the clean world of resonating orbits for the
dirty world of turbulent, magnetized, radiating discs with
unpleasant boundary conditions, but QPOs occur in such world.
Abramowicz believes that QPOs are the Rosetta stone for
understanding black-hole accretion. Not so. If one had to
(over)use\footnote{The road to the theorist's hell is paved with
Rosetta stones} the Rosetta-stone analogy, QPOs would be just one of
the texts on this stone. Let's hope it is the Greek one. All in all,
these publications are not so bad: imagine a volume devoted to the
beat-frequency model. At least the epicyclic resonance model is
still alive.

The authors of the papers  deal only with neutron star and
black-hole QPOs. The abundant QPOs observed in CVs are only
mentioned en passant and no special attention is paid to them.
Probably because, not being (sufficiently) relativist, they are
considered boring. In view of the recently published article on the
subject \cite{klab} such an attitude is rather surprising.

\subsection*{Observations}

The four contributions in this category have been written by some of
the top observers of X-ray binaries and they form a very good (too
good maybe) background for the theoretical papers. van der Klis, as usual, gives a clear and sober
review of the QPO phenomenon. One wishes theorists paid more
attention to what he has to say about black hole hfQPOs: \textsl{The
phenomenon is weak and transient so that observations are difficult,
and discrepant frequencies occur as well, so it can not be excluded
that these properties of approximately constant frequency and
small-integer ratios would be contradicted by work at better signal
to noise.} Being a loyal participant he adds: \textsl{In the
remainder I will assume these properties are robust.}

A usual in QPO research it is difficult to get used to the
terminology and classification. It took some time to make sense of
\textsl{atolls}, \textsl{bananas} and \textsl{z}-\textsl{tracks}
(and sources!) and now we encounter the challenge of the X-ray
states of Black Hole Binaries. Not surprisingly Remillard is using
the classification defined in his monumental work with McClintock
\cite{mcrm}. We have therefore the \textsl{thermal}, \textsl{hard}
and \textsl{SPL} states. One might be slightly worried not seeing
the \textsl{thermal dominant (TD) state} \cite{mcrm} but fortunately
we are told that the thermal state is the \textsl{formerly
``high/soft" state"}, so \textsl{TD = thermal}. In any case the real
drama begins when one wishes to see what other specialists have to
say about the subject, e.g. Belloni (2005). There we find a
different classification into: an \textsl{LS} (Low/hard state), an
\textsl{HIMS} (High Intermediate State), a \textsl{SIMS} and an
\textsl{HS} (High/Soft state). It seems that \textsl{HS=TD} and
\textsl{LS=hard} but in the two other cases relations are not clear.
This is not surprising because Belloni defines his states by the
transition properties and not by the state properties. In addition
Belloni (2005) classifies low frequency QPOs into A, B and C types,
whereas Remillard uses quantities $a$ and $r$, the rms amplitude
and power (note that it was Remillard who introduced type C QPOs).
Both approaches have their merits and one can understand why they
were introduced but they make life really difficult for people
trying to understand the physics of accretion flows. I am surprised
that Abramowicz complains only about the confusion introduced by
numerical simulations and not about the impenetrable jungle of X-ray
states and QPO terminology. I suspect he has given up on reading on
this subject.

However, Remillard convincingly shows that hfQPOs appear in the
SPL state and shows very interesting relations between the presence
of $2\nu_0$ and the $3\nu_0$ frequencies and the state of the system
as described by the disc flux and the power-law flux. As far as I
can tell this is ignored by the epicyclic theorists but this could
be the second text of the Rosetta stone. It is also a major
difficulty for the epicyclic resonance model. Since the SPL state is
characterized by a strong Comptonised component in the X-ray flux,
it is difficult to see how the flux modulation at the vertical
epicyclic frequency by gravitational-lensing could survive in such
an environment.

This brings me to the contribution by Barret and collaborators.
Recently Barret with a different (but intersecting) set of
collaborators \cite{barretal} made a fundamental discovery by
showing that the lower frequency kHzQPO in the neutron-star binary
4U 1608-52 is a highly coherent signal that can keep $Q\approx 200$
for $\sim 0.1$ s. They also found that the higher frequency kHzQPO
is fainter than its lower frequency counterpart and has lower $Q$.
Barret et al. (2005) showed very convincingly that no proposed QPO
model can account for such highly coherent oscillations. They can
all be rejected except for the ERM but only because the two resonant
oscillators have not yet been identified. In particular, they
rejected the modified beat-frequency model of Miller et al. (1998).
In Barret et al. another puzzling phenomenon is presented. They
found in three neutron-star binaries (including 4U 1608-52) showing
high-Q lower kHzQPOs that the coherence increases with frequency to
a maximum ($\sim 800$ Hz) after which it rapidly drops and QPOs
disappear. To me it looks like an effect related to the forcing
mechanism. Barret et al. link their observations to the ISCO basing
their claim on the Miller et al. (1998) model. There is half a
paragraph trying to explain (I think) how the model rejected in a
previous paper can be rejuvenated (or rather resuscitated) and used
to interpret the present observations. I read this part of the paper
several times and failed to understand its meaning. I had no problem
understanding the reasoning rejecting Miller et al. (1998).

In any case I also fail to understand why the Barret et al. (2005)
discovery of the high coherence of QPOs was not the central point of
the Nordita workdays.  It is easy to miss a \textit{Mane, Mane,
Tekel, Uphar'sin} when looking for a Rosetta stone.

The main result of the excellent article on neutron-star boundary
layers Gilfanov is that the kHzQPOs appear to have the same origin
as aperiodic  and quasiperiodic variability at lower frequency. It
seems to be clear that the msec flux modulations originate on the
surface of the neutron star. Nota bene, I am surprised that the
remarkable and extremely relevant discovery of the universal
rms-flux correlation (Uttley 2004; Uttley et al. 2005) is not
mentioned in this context. Gilfanov
point out that the kHz clock could still be in the disc.

\subsection*{Disc simulations}

It is known that in stars some multimode pulsations may arise from
stochastic excitation by turbulent convection (see e.g. Dziembowski
2005). It is therefore legitimate to expect that in turbulent discs
similar effects could be found. Brandenburg presents very
interesting results obtained in the framework of the shearing-box
approximation of accretion disc structure. He obtains what he calls
stochastic excitation of epicycles. In his model the radial
epicyclic frequency is equal to the Keplerian frequency and the
vertical epicyclic frequency is not equal (or comparable) to the
p-mode frequency so it is not clear how close his results are to
what is happening in full-scale discs. But they are promising.
Another result concerning dissipation in discs requires more
investigation. According to Brandenburg in MRI discs most of the
dissipation occurs in the corona, whereas in the forced hydrodynamic
case most of the dissipation occurs near the midplane. He claims
that his result, obtained in the isothermal case, has been shown
also for radiating discs. The disc model in question, however, was
radiation-pressure dominated while gas-pressure dominated models
\cite{millstone} do not seem to confirm the claim that MRI discs
release most of the energy in the corona.

\subsection*{The epicyclic resonance model}

The eleven contributions to the epicyclic resonance model  contain two
general articles by the founders; the other papers on different
aspects of the model were written (except for the last contribution)
by younger members of the ERM team.  All these
contributions are very well written, clear and to the point. I was
really impressed by their quality. They contain all one needs to
know about the ERM. As far as I know they were written by the
authors whose names appear explicitly on the paper and since they
are very careful in acknowledging other people's contributions I
recommend removing the ``et al.'s" which give the impression that
the texts were written by a sect, or that they form a sort of
Norditan Creed. Fortunately this is not the impression one gets
reading the articles. They are professional, open to alternatives,
pointing out difficulties etc.

Of particular quality in this respect in the contribution by Paola
Rebusco. She presents the problem in a very clear way and
carefully chooses the (difficult) questions still to be answered.
Ji\v{r}\'{\i} Hor\'ak contributes two interesting articles. The
first discusses the 3:2 autoparametric resonance in the general
framework of conservative systems and shows that the amplitude and
frequency of the oscillations should be periodically modulated - a
result that might relate hfQPOs to lower frequency QPOs. The second
paper tries to explain the QPO modulations in neutron-star binaries
by a mechanism proposed by Paczy\'nski. It is not clear if such a
mechanism could achieve the high quality factors observed by Barret
et al. (2005) or how it relates to the oscillation forced by the
spinning neutron-star magnetic field. Three contributions deal with
various aspects of oscillating tori. Eva {\v S}r{\' a}mkov{\' a}
presents the preliminary results of her research on eigenvalues and
eigenfrequencies of slightly non-slender tori. She includes in her
paper a figure showing a transient torus appearing in a 3D
simulation of an accretion flow -- a rather touching testimony to the
ERM-team reliance on this elusive structures. William Lee uses SPH
simulations to study the response of a slender torus to external
periodic forcing. The results are a very interesting illustration of
the essentially nonlinear character of the coupling between the
radial and vertical modes (coupling through the sub-harmonic of the
perturbation: $1/2\nu_0$) and the rather fascinating phenomenon of
mode locking for a drifting torus. This can be relevant to the drift
of QPO frequencies observed in neutron-star binaries. Since his
contribution is devoted to these systems, mentioning ``stellar-mass
black holes" in the abstract is a bit misleading. Michal Bursa
expertly attacks the problem crucial for the ERM applied to black
holes: how to produce \textsl{two} modulations of the X-ray flux. By
using a toy model consisting of an optically thin, bremsstrahlung
emitting, oscillating slender torus he shows that strong-gravity
relativistic effects may produce the desired result. How would
things look in the case of an optically thick disc surrounded by a
comptonizing cloud is (probably) a different story. The last three
contributions deal with some aspects of hfQPO observations. Tomek
Bulik reanalysis the somewhat controversial issue of the Sco X-1
kHzQPO clustering around the value corresponding to the frequency
ratio of 2/3. His skillful analysis shows that the clustering is
real.
Gabriel T{\"o}r{\"o}k has been entrusted with the somehow irksome
task of linking microquasar QPOs with those observed in Sgr A$^*$
and AGNs. Since the last category forms an empty set he could just
discuss why such observations would be important. Unfortunately the
prospect of detecting QPOs from AGNs is rather remote
\cite{vauguttl}. His valiant attempt to discuss determining
black-hole spin from hfQPOs was hindered by the uncertainties in
both data and models. But his is a very good short review of the
problem.

Because they are a general introduction to an unfinished
construction, the contributions by the founders are less
interesting. Abramowicz gives a general introduction to the
subject of accretion onto compact objects. In his (entirely
justified) efforts to rehabilitate his and collaborators' (to whom I
belong) fundamental contributions to the concept of ADAF, Abramowicz
went too far: he antedated the relevant Abramowicz et al. paper by
ten years and did not insert the Narayan \& Yi article into the
references. I think also that his claim that accretion theory today
experiences  a period of confusion caused by supercomputer
simulations is exaggerated. The confusion is caused by (some)
astrophysicists hastily trying to apply to real objects whatever
comes out of the computer and not by the physicists making these
very impressive simulations. People who are confused should read the
excellent article by  
 Balbus (2005) -- a real guide for the
perplexed. However, Eq.~(2) Abramowicz can create confusion since
it asserts that the radial epicyclic frequency is \textsl{larger}
than the vertical one. Luckily there is his Fig.~2 to sober us up.
Klu\'zniak with his usual  charming intellectual incisiveness
describes his personal road to ERM. He is convinced that after
trying various roads which led nowhere, he finally chose the right
one. He knows it is uphill and very steep. But never send to know
for whom the disc tolls; it tolls for him. I wish him luck.

\acknowledgements  I am grateful to Marek Abramowicz for inviting me
to write this report and to G\"unther R\"udiger for accepting this
risky idea.

\end{document}